\documentclass[german,english]{bvm2018}

\usepackage{amsmath}
\usepackage{gensymb}
\usepackage[tophrase={~to~},binary-units=true,separate-uncertainty = true,multi-part-units=single]{siunitx}
\usepackage{multirow}
\usepackage{booktabs}
\usepackage{pgfplots}
\pgfplotsset{compat=newest}

\usepackage{todonotes}
\usepackage{xcolor}
\usepackage{blindtext}

\title{Effects of Tissue Material Properties on X-Ray Image, Scatter and Patient Dose Determined using Monte Carlo Simulations}
\titlerunning{Effect of Tissue Material Properties on Monte Carlo Simulation}

\author{Philipp~Roser$^{1,3}$, Annette~Birkhold$^2$, Xia~Zhong$^1$, Elizaveta~Stepina$^2$, Markus~Kowarschik$^2$, Rebecca~Fahrig$^2$, Andreas~Maier$^{1,3}$}

\authorrunning{Roser et al.}

\institute{%
 $^1$Pattern Recognition Lab, FAU Erlangen-N\"urnberg\\
 $^2$Siemens Healthcare GmbH, Forchheim Germany\\
 $^3$Erlangen Graduate School in Advanced Optical Technologies (SAOT)}

\email{philipp.roser@fau.de}

\begin{document}

%
\selectlanguage{english}

\maketitle

\begin{abstract}
With increasing patient and staff X-ray radiation awareness, many efforts have been made to develop accurate patient dose estimation methods. 
To date, Monte Carlo (MC) simulations are considered golden standard to simulate the interaction of X-ray radiation with matter. However,  sensitivity of MC simulation results to variations in the experimental or clinical setup of image guided interventional procedures are only limited studied. In particular, the impact of  patient material compositions  is poorly investigated. This is mainly due to the fact, that these methods are commonly validated in phantom studies utilizing a single anthropomorphic phantom. 
In this study, we therefore investigate the impact of patient material parameters mapping on the outcome of MC X-ray dose simulations. A computation phantom geometry is constructed and three different commonly used material composition mappings are applied.
We used the MC toolkit Geant4 to simulate X-ray radiation in an interventional setup and compared the differences in dose deposition, scatter distributions and resulting X-ray images.
The evaluation shows a discrepancy between different material composition mapping up to \SI{20}{\percent} concerning directly irradiated organs.
These results highlight the need for standardization of material composition mapping for MC simulations in a clinical setup.

\end{abstract}

\section{Introduction}
Over the last years, the amount of X-ray guided diagnostic and interventional procedures has increased steadily, raising the awareness of dose-induced deterministic and stochastic risks for the patient as well as the treating medical staff.
Therefore, efforts are made to determine and visualize the distribution of absorbed dose and scattered radiation in the context of the interventional suite and hybrid operating room using Monte Carlo (MC) methods~\cite{Rodas:15}.
Recently, MC simulation of photon transport gained additional boost with deep convolutional neural networks being established to be state of the art in most X-ray imaging classification and regression tasks, such as landmark detection or segmentation.
With novel architectures emerging on a daily basis, the demand for diverse training and testing data intensifies. 
Since medical data is treated sensitively, there is a constant lack of sufficient data. 
Although efforts are made to build open source databases, there exist prominent problems, such as scatter reduction~\cite{Maier:18}, hindering the collection of accurate ground truth data without imitating existing solutions, such as anti-scatter grids.
Therefore, realistic simulation of these problems has become a fundamental step to build learning solutions to real-world problems.
However, to push deep learning from research to clinical application, the training data must be valid to a certain measure.
There is, however, a multitude of parameters affecting the outcome of MC simulations in an unintuitive way, such as modeling the energy spectrum or biasing the particle source.
To obtain valid and realistic results, it is mandatory to be aware of all sources of uncertainty concerning modeling the clinical setup.
In this study, the impact of variations in the tissue material properties on resulting X-ray image, scattered radiation and patient dose are determined using Monte Carlo simulation. 

\section{Materials and Methods}

\subsubsection{Phantom Model Geometry and Material Parameters}
To study the effect on material composition mapping, we use the geometry of the voxel phantom Golem provided by the Institute for Radiation Protection\,\footnote{www.helmholtz-muenchen.de/iss/index.html}.
The Golem phantom consists of \num{220} slices with \num{256 x 256} voxels each, ranging from the vertex down to the toes of a normally shaped, \SI{176}{\cm} adult male.
It is segmented into \num{122} organ and tissue labels.
Three different, voxel-wise material composition mappings are used to assign material properties to the associated labels for MC simulation.
A material is defined by its volumetric mass density and the fraction of mass of elementary components.
Two material composition mappings reference the commonly used anthropomorphic dosimetry phantoms RANDO (Alderson\,\footnote{www.rsdphantoms.com/rt\_art.htm}) and CIRS (ATOM\,\footnote{www.cirsinc.com/products/all/33/atom-dosimetry-verification-phantoms/}), respectively. The Alderson mapping \emph{AM1} includes real bone (cortical) and an approximation of the lungs besides a mixture to represent soft tissue as the main component of the human body. The Atom mapping \emph{AM2} includes bone, soft and lung (inhale) equivalent tissues.
The third material mapping serves as reference mapping \emph{RM}  and is modeled to resemble a living adult male, following the material specifications proposed by the International Commission on Radiological Protection (ICRP)\,\footnote{www.icrp.org} standard.
It comprises adipose, soft, skin, brain, bone (cortical), muscle and lung (inhale) tissue. 

\subsubsection{Detector Model}
The simulated \SI{320 x 237.5}{\mm} flat panel detector has a resolution of \num{256 x 190} pixels.
To reduce variance, it consists of Cesium-Iodide with a \SI{20}{\mm} thickness to absorb all incoming photons.
We consider the detector as an ideal detector with a linear detector response curve, no electron noise or defect pixels. 
No processing is applied to the resulting image from the detector.

\subsubsection{Simulation of Experimental Setup}
\label{subsec:experiment-configuration}
The simulation is implemented in the general purpose MC toolkit Geant4 \cite{Agostinelli:03}, which offers a high degree of customizability and flexibility allowing for arbitrary experiment configuration and quantity scoring.
Furthermore, Geant4 provides an interface to materials as defined by the ICRP, alongside arbitrary material compositions.

The phantom is centered in the origin of the world coordinate system, the particle source is placed in \SI{800}{\mm} distance ante-posterior to the phantom, such that the prostate lies approximately in the center of the emitted X-ray beam.
The particle source is circularly shaped with a radius of \SI{0.3}{\mm} and collimated resulting in \SI{7.6}{\degree} for both aperture angles.
Emitted photon vertices are sampled using cosine-weighting to obtain homogeneous fluence with respect to a sphere surface.
The underlying energy spectrum of the photon shower is modeled considering a tungsten anode, \SI{70}{\kilo\volt} peak voltage and \SI{2.7}{\mm} Aluminum self-filtration using Boone's algorithm~\cite{Boone:97}.
The flat panel detector is placed in \SI{1300}{\mm} distance to the photon source perpendicular to the central X-ray direction.
Particle interactions that may occur at the given energy spectrum are considered, including the photo electric effect, Rayleigh scattering and Compton scattering for photons and ionization and Bremsstrahlung for electrons. 
All processes are modeled adhering to the Livermore model for low energy physics~\cite{Cullen:97}.
Primary photons and secondary particles are tracked until their associated kinetic energy in consumed completely to satisfy energy preservation and assure accurate results.

To obtain stable dose and scatter distributions \num{9e8} primary photons are emitted, for X-ray image generation \num{51e8}, respectively.
Dose distributions are scored with respect to the dose $D$ absorbed by each voxel measured in \si{\gray}. 
To quantify scatter distributions and X-ray images, the incident radiant energy $R$ in \si{\joule} is tracked.
The simulation is carried out in batches of \num{e8} primaries in order to bring variance to the initial random seed and to split the computation to several nodes of the high performance computing (HPC) cluster.
Each batch computation lasts on average \SI{3.5}{\hour}, however multiple batches are processed in parallel.
The resulting dose distributions have the same resolution as the associated phantom volumes.
To score the scatter distributions, a \SI{8}{\m^3} volume comprising \num{100 x 100 x 100} isotropic voxels is placed surrounding the phantom and  material parameters of air defined by ICRP are applied. No interventional table is considered. We performed simulations using aforementioned configurations for each mapping. The simulation result employing \emph{RM} are considered as base line, results of \emph{AM1} and \emph{AM2} are compared to this reference.

\section{Results}
\subsubsection{Scatter and Dose Distributions}
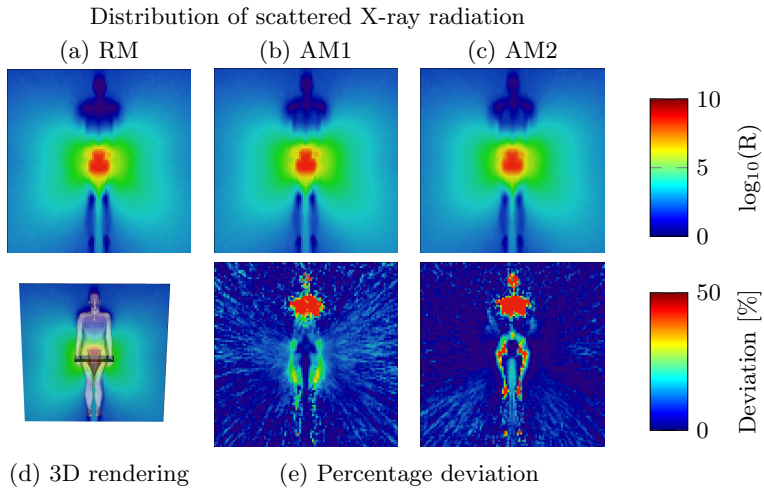
\begin{figure}[ptb]
    \centering
    \begin{tabular}{@{\extracolsep{6pt}} c c c l@{}}
         \multicolumn{3}{c}{Distribution of scattered X-ray radiation} \\
         (a) RM & (b) AM1 & (c) AM2  \\
         \includegraphics[width=0.2\textwidth]{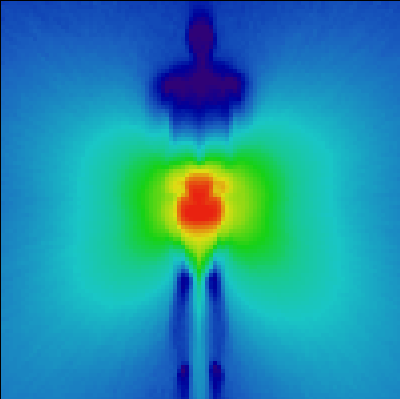} 
         & \includegraphics[width=0.2\textwidth]{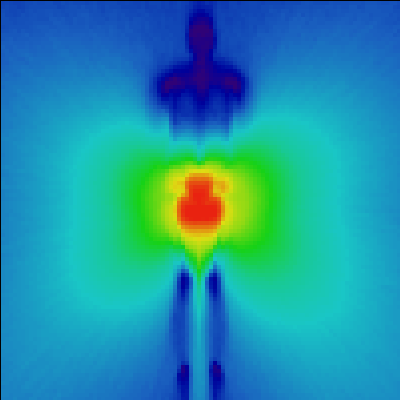} 
         & \includegraphics[width=0.2\textwidth]{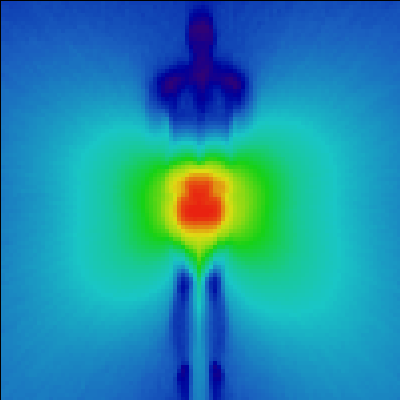} 
         & \begin{tikzpicture}
				\begin{axis}[
    				hide axis,
    				scale only axis,
    				height=0pt,
    				width=0pt,
    				colormap/jet,
    				colorbar,
    				point meta min = 0,
    				point meta max = 10,
    				colorbar style={
    				    height=0.15\textwidth, 
    				    align=center,
    				    ytick={0, 5, 10},
                        ylabel=log$_{10}$(R),
                        ylabel near ticks, 
                        yticklabel pos=right
                    }  
    			]
    				\addplot [draw=none] coordinates {(0,0)};
				\end{axis}
			\end{tikzpicture} \\ 
         \includegraphics[width=0.2\textwidth]{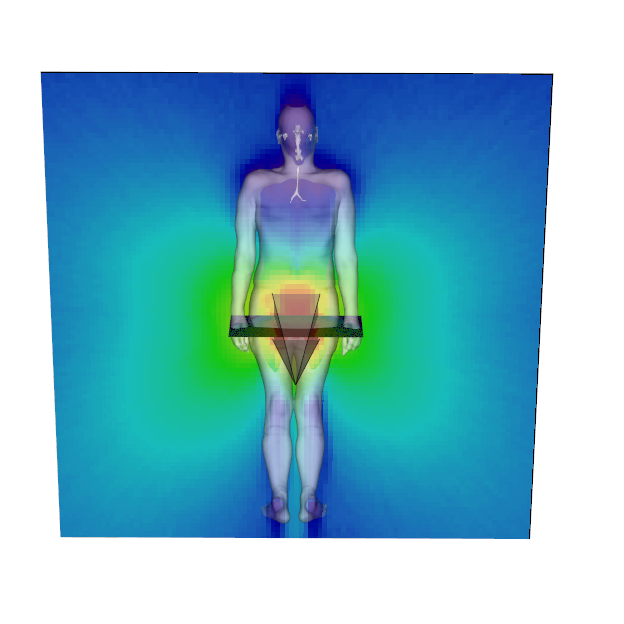} 
         & \includegraphics[width=0.2\textwidth]{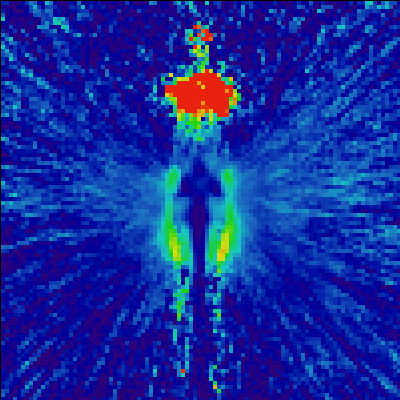} 
         & \includegraphics[width=0.2\textwidth]{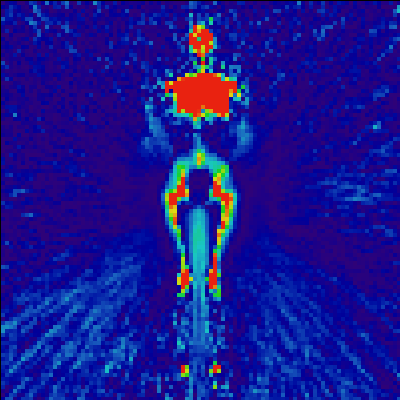} 
         & \begin{tikzpicture}
				\begin{axis}[
    				hide axis,
    				scale only axis,
    				height=0pt,
    				width=0pt,
    				colormap/jet,
    				colorbar,
    				point meta min = 0,
    				point meta max = 50,
    				colorbar style={
    				    height=0.15\textwidth, 
    				    align=center,
    				    ytick={0, 50},
                        ylabel=Deviation [\si{\percent}],
                        ylabel near ticks, 
                        yticklabel pos=right
                    }  
    			]
    				\addplot [draw=none] coordinates {(0,0)};
				\end{axis}
			\end{tikzpicture} \\  
         (d) 3D rendering & \multicolumn{2}{c}{(e) Percentage deviation} \\
    \end{tabular}
    \caption{Coronal view of the scatter distributions associated with each material composition mapping.  (e) Corresponding percentage deviation maps with respect to RM. (d) Spatial relationship between scatter maps and phantom. Scatter distributions are shown in logarithmic domain.  The identifier $R$ refers to the radiant energy entering a voxel in \si{\joule}.}
    \label{fig:scatter-dist}
\end{figure}

\begin{figure}[ptb]
    \centering
    \begin{tabular}{@{\extracolsep{6pt}} c c c l@{}}
         \multicolumn{3}{c}{Distribution of deposited X-ray dose} \\
         (a) RM & (b) AM1 & (c) AM2  \\
         \includegraphics[width=0.2\textwidth]{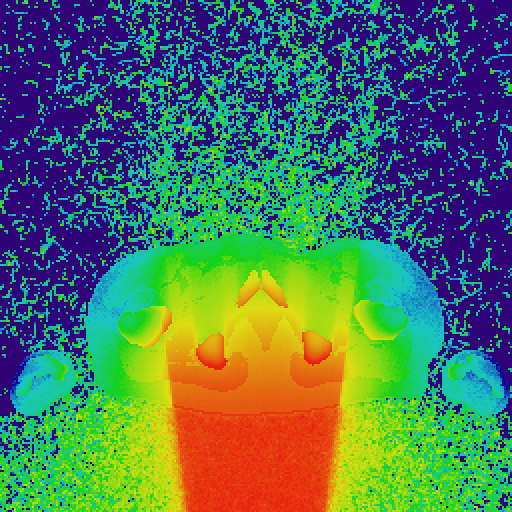} 
         & \includegraphics[width=0.2\textwidth]{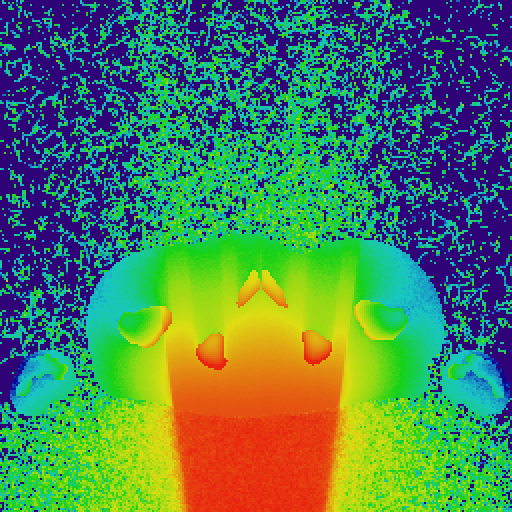} 
         & \includegraphics[width=0.2\textwidth]{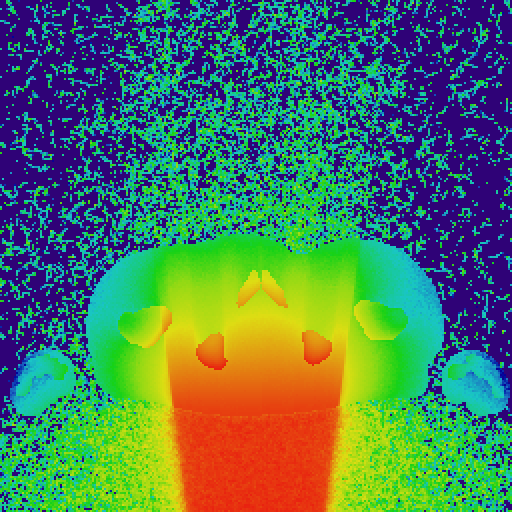} 
         & \begin{tikzpicture}
				\begin{axis}[
    				hide axis,
    				scale only axis,
    				height=0pt,
    				width=0pt,
    				colormap/jet,
    				colorbar,
    				point meta min = -23,
    				point meta max = -17,
    				colorbar style={
    				    height=0.15\textwidth, 
    				    align=center,
    				    ytick={-23, -20, -17},
                        ylabel=log$_{10}$(D),
                        ylabel near ticks, 
                        yticklabel pos=right
                    }  
    			]
    				\addplot [draw=none] coordinates {(0,0)};
				\end{axis}
			\end{tikzpicture} \\ 
         \includegraphics[width=0.2\textwidth]{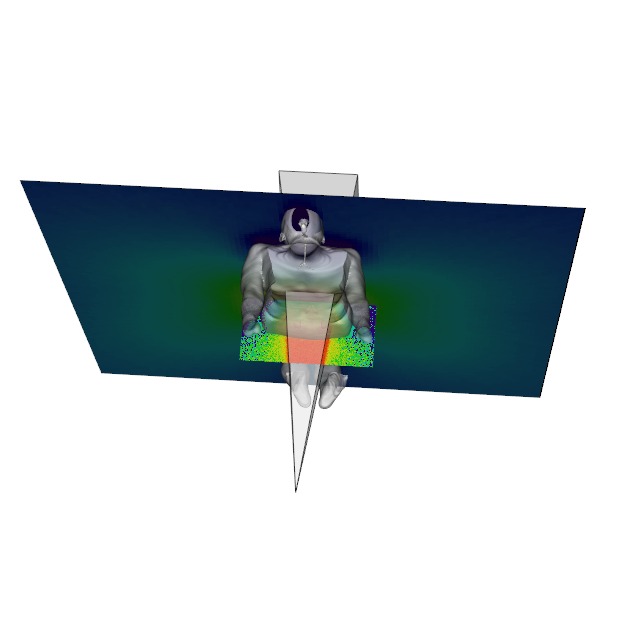} 
         & \includegraphics[width=0.2\textwidth]{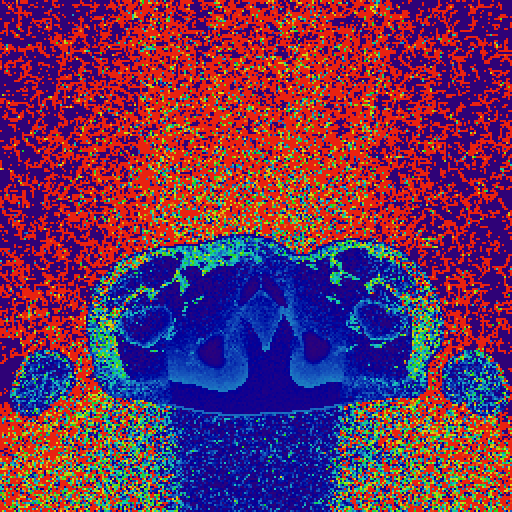} 
         & \includegraphics[width=0.2\textwidth]{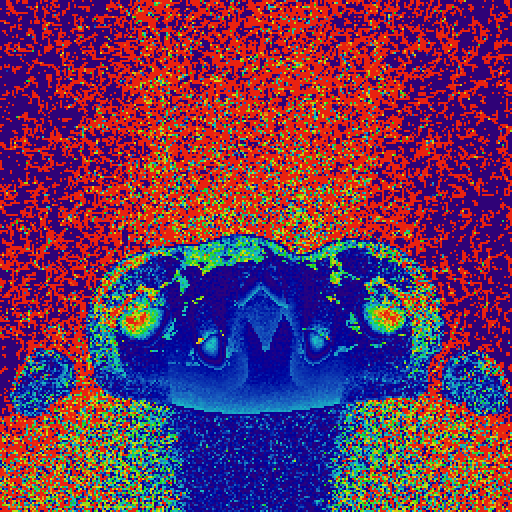} 
         & \begin{tikzpicture}
				\begin{axis}[
    				hide axis,
    				scale only axis,
    				height=0pt,
    				width=0pt,
    				colormap/jet,
    				colorbar,
    				point meta min = 0,
    				point meta max = 50,
    				colorbar style={
    				    height=0.15\textwidth, 
    				    align=center,
    				    ytick={0, 50},
                        ylabel=Deviation [\si{\percent}],
                        ylabel near ticks, 
                        yticklabel pos=right
                    }  
    			]
    				\addplot [draw=none] coordinates {(0,0)};
				\end{axis}
			\end{tikzpicture} \\ 
         (d) 3D rendering & \multicolumn{2}{c}{(e) Percentage deviation} \\
    \end{tabular}
    \caption{(a)-(c) Axial view of the deposited X-ray dose distribution associated with each set of material properties. (e) Corresponding percentage deviation maps with respect to RM. (d) Spatial relation between  dose maps and phantom. The dose distributions are shown in the logarithmic domain. The identifier $D$ refers to the dose absorbed by a voxel in \si{\gray}.}
    \label{fig:dose-dist}
\end{figure}
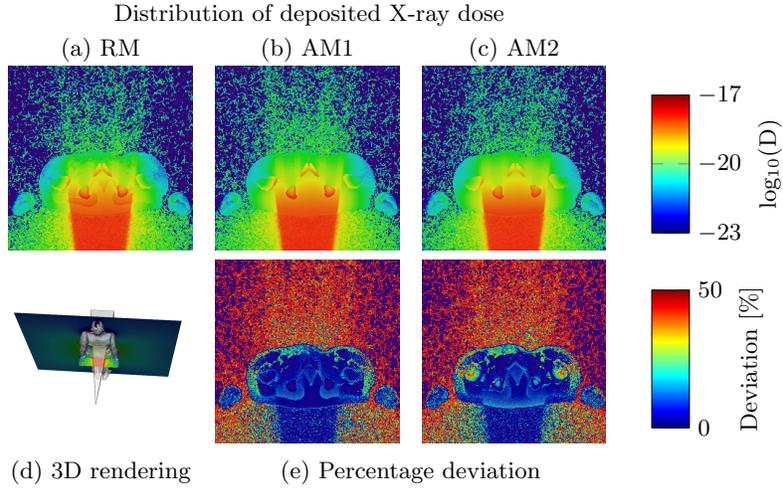

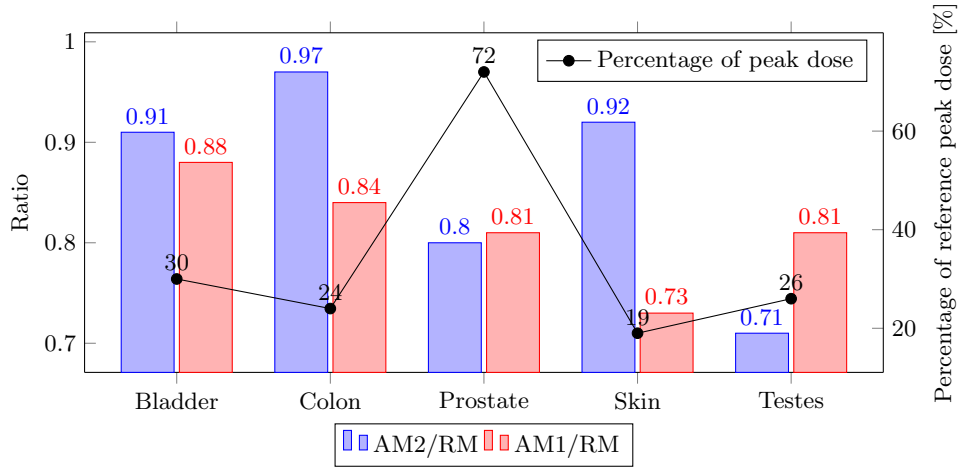
\begin{figure}[ptb]
    \centering
    \begin{tikzpicture}
        \begin{axis}[
            axis y line*=left,
            ybar,
            bar width = 20pt,
            width=\textwidth,
            height=0.5\textwidth,
            enlargelimits=0.15,
            legend style={at={(0.5,-0.15)},
              anchor=north,legend columns=-1},
            ylabel={Ratio},
            symbolic x coords={Bladder, Colon, Prostate, Skin, Testes},
            xtick=data,
            nodes near coords,
            nodes near coords align={vertical},
            ]
            \addplot coordinates {(Bladder, 0.91) (Colon, 0.97) (Prostate, 0.80) (Skin, 0.92) (Testes, 0.71) };
            \label{am2rm}
            \addlegendentry{AM2/RM}
            \addplot coordinates {(Bladder, 0.88) (Colon, 0.84) (Prostate, 0.81) (Skin, 0.73) (Testes, 0.81) }; 
            \label{am1rm}
            \addlegendentry{AM1/RM}
        \end{axis}
        \begin{axis}[
            axis y line*=right,
            axis x line = none,
            width=\textwidth,
            height=0.5\textwidth,
            enlargelimits=0.15,
            symbolic x coords={Bladder, Colon, Prostate, Skin, Testes},
            xtick=data,
            nodes near coords,
            nodes near coords align={vertical},
            ylabel={Percentage of reference peak dose [\si{\percent}]}
            ]
            \addplot[mark=*] coordinates { (Bladder, 30) (Colon, 24) (Prostate, 72) (Skin, 19) (Testes, 26) };
            \addlegendentry{Percentage of peak dose}
        \end{axis}
    \end{tikzpicture}
    \caption{Ratios of total organ dose between different material composition mappings (AM1, AM2) and the reference (RM) for five directly irradiated organs.
    The absolute organ doses in \si{\gray} for the RM are given by the black plot. }
    \label{fig:organ-dose}
\end{figure}

Fig.~\ref{fig:scatter-dist}a-c show the distributions of scattered radiation in the patient environment ($\log_{10}$; coronal slices) using the three material mappings.
The deviation maps of the percentage difference to \emph{RM} for \emph{AM1}  and \emph{AM2}  are depicted in fig.~\ref{fig:scatter-dist}e.
Distribution of scattered radiation in both \emph{AM} simulations shows high overall agreement with the \emph{RM} results; however, concerning specific regions deviations of \SIrange{20}{50}{\percent} were determined.
Fig.~\ref{fig:dose-dist} shows an axial slice of the phantom dose distribution simulation results for \emph{RM} (a), \emph{AM1} (b) and \emph{AM2} (c).
Deviation maps of the dose distribution for \emph{AM1}  and \emph{AM2} are depicted in fig.~\ref{fig:dose-dist}e and show similar deviations from the reference as the scattered radiation.
For a set of directly irradiated organs (bladder, colon, prostate, skin, testes) the total dose was determined.
Fig.~\ref{fig:organ-dose} shows the dose ratio between \emph{AM}s and \emph{RM} for these dose sensitive organs.
Correlating the \emph{AM}s to the \emph{RM}, introduces a deviation of up to \SI{20}{\percent} for the prostate at a reference dose of \SI{72}{\percent} of the peak dose measured.
For organs within \SIrange{19}{30}{\percent} of the peak dose, a deviation of \SIrange{3}{29}{\percent} can be observed.

\subsubsection{X-Ray Images}
\begin{figure}[ptb]
    \centering
    \begin{tabular}{@{\extracolsep{6pt}} c c c l@{}}
         \multicolumn{3}{c}{Dose distribution} \\
         (a) RM & (b) AM1 & (c) AM2  \\
         \includegraphics[width=0.25\textwidth]{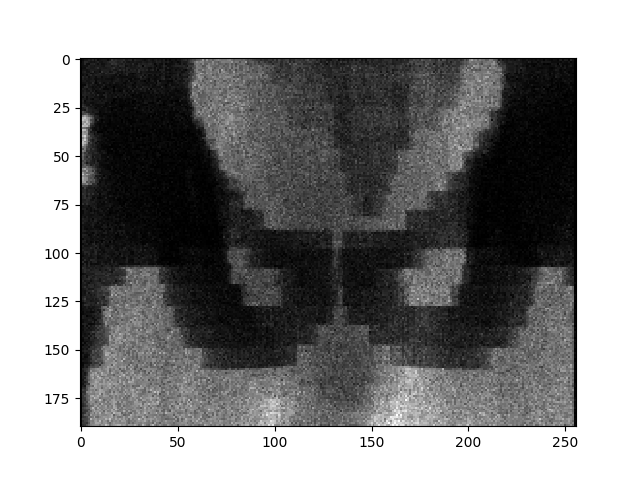} 
         & \includegraphics[width=0.25\textwidth]{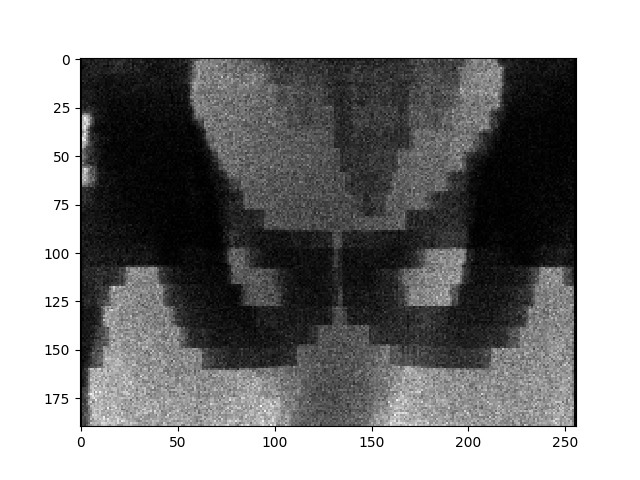} 
         & \includegraphics[width=0.25\textwidth]{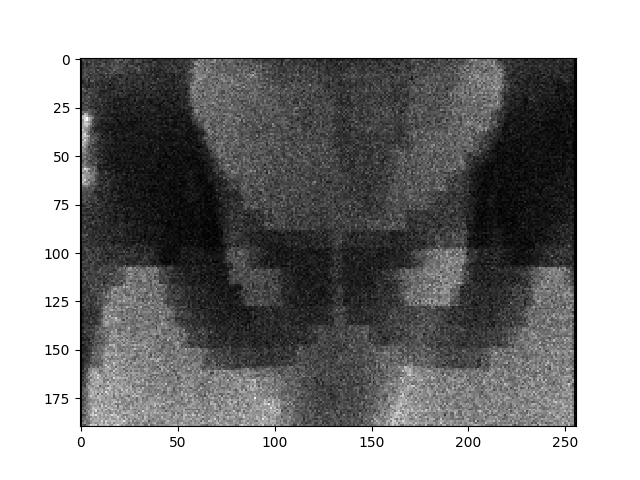} 
         & \begin{tikzpicture}
				\begin{axis}[
    				hide axis,
    				scale only axis,
    				height=0pt,
    				width=0pt,
    				colormap/blackwhite,
    				colorbar,
    				point meta min = 0,
    				point meta max = 5,
    				colorbar style={
    				    height=0.1\textwidth, 
    				    align=center,
    				    ytick={0, 5},
                        ylabel=Energy [\si{\joule}],
                        ylabel near ticks, 
                        yticklabel pos=right
                    }  
    			]
    				\addplot [draw=none] coordinates {(0,0)};
				\end{axis}
			\end{tikzpicture} \\ 
          & \includegraphics[width=0.25\textwidth]{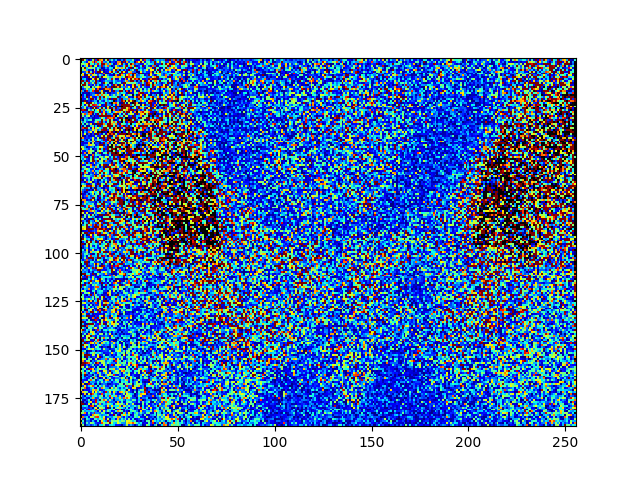} 
          & \includegraphics[width=0.25\textwidth]{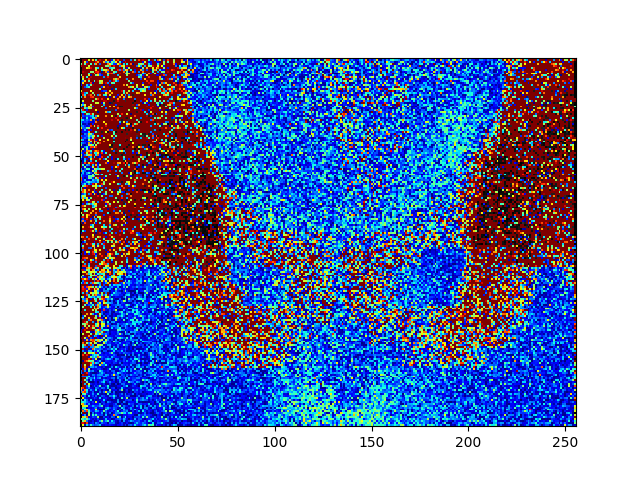} 
          & \begin{tikzpicture}
				\begin{axis}[
    				hide axis,
    				scale only axis,
    				height=0pt,
    				width=0pt,
    				colormap/jet,
    				colorbar,
    				point meta min = 0,
    				point meta max = 100,
    				colorbar style={
    				    height=0.1\textwidth, 
    				    align=center,
    				    ytick={0, 100},
                        ylabel=Deviation [\si{\percent}],
                        ylabel near ticks, 
                        yticklabel pos=right
                    }  
    			]
    				\addplot [draw=none] coordinates {(0,0)};
				\end{axis}
			\end{tikzpicture} \\ 
          & \multicolumn{2}{c}{(d) Percentage deviation} \\
    \end{tabular}
    \caption{Primary photon contribution to X-ray images generated with respect to \num{51e8} primary particles. No processing is applied, the raw radiant energy incident at the detector is tracked.}
    \label{fig:x-ray}
\end{figure}
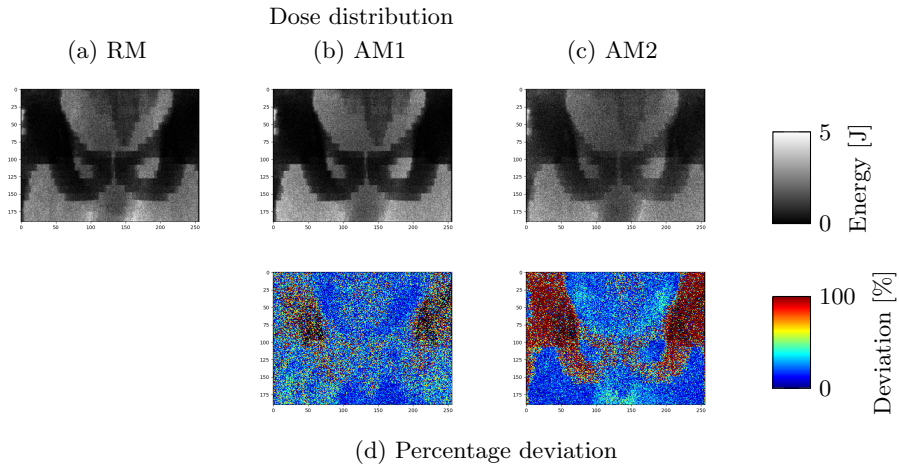

Fig.~\ref{fig:x-ray} shows the detector image results of the simulation and associated deviation maps. 
Although the images are similar in general, the deviation maps disclose major differences concerning all tissue types. 
Future studies may evaluate if these differences are in a diagnostic relevant range.

\section{Summary}
This study highlights variances in MC simulation results when using different material composition mapping for the same phantom geometry. We showed, that the material composition mapping affects X-ray dose, scatter as well as created image to a certain extent. Therefore, for quantitative analysis and comparison between experimental and simulation studies these variances have to be considered. A more detailed standardization of material parameters might be needed. This need for standardization is further emphasized as MC simulations are potentially used to generate training data for deep learning methods.
~\\
\textbf{Disclaimer:} The concepts and information presented in this paper are based on research and are not commercially available.

\bibliographystyle{bvm2018}

\bibliography{0000}
\end{document}